\documentclass[twocolumn,showpacs,prl,superscriptaddress,floatfix]{revtex4}

\usepackage{color} 
\usepackage{tabularx} 
\usepackage{epsfig}
\usepackage{amsmath} 
\usepackage{amssymb} 
\usepackage{graphicx}
\usepackage{wasysym}
\usepackage{times}

\newcommand{\vect}[1]{\boldsymbol{\mathrm{#1}}}

\begin{document}

\title{Error Threshold for Color Codes and Random 3-Body Ising Models}

\author{Helmut G.~Katzgraber} 
\affiliation {Theoretische Physik, ETH Zurich, CH-8093 Zurich, Switzerland}
\affiliation {Department of Physics, Texas A\&M University, College Station,
Texas 77843-4242, USA}

\author{H.~Bombin}
\affiliation{Department of Physics, Massachusetts Institute of Technology, 
Cambridge, Massachusetts 02139, USA}

\author{M.~A.~Martin-Delgado}
\affiliation{Departamento de F{\'i}sica Te{\'o}rica I, Universidad 
Complutense, 28040 Madrid, Spain}

\date{\today}

\begin{abstract}

We study the error threshold of color codes, a class of topological
quantum codes that allow a direct implementation of quantum Clifford
gates suitable for entanglement distillation, teleportation and
fault-tolerant quantum computation. We map the error-correction
process onto a statistical mechanical random 3-body Ising model and
study its phase diagram via Monte Carlo simulations.  The obtained
error threshold of $p_c = 0.109(2)$ is very close to that of Kitaev's
toric code, showing that enhanced computational capabilities do not
necessarily imply lower resistance to noise.

\end{abstract}

\pacs{03.67.Lx, 75.40.Mg, 03.67.Pp, 75.50.Lk}


\maketitle

Protecting quantum states from external noise and errors is central
for the future of quantum information technology.  Because interaction
with the environment is unavoidable, active quantum error-correction
techniques based on quantum codes have been devised to restore
the damaged quantum states from errors caused by decoherence
\cite{shor:95,steane:96}. These approaches are, in general, cumbersome
and require many additional quantum bits, thus making the system
more error prone.  An imaginative and fruitful approach to quantum
protection is to exploit topological properties of a system, e.g.,
by using the nontrivial topology of a surface to encode quantum
states at the logical level \cite{kitaev:03}.  Topology is thus
considered as a resource, much like entanglement is a resource for
quantum information tasks. Topological quantum computation is the
combination of these two resources with the aim of winning the battle
against decoherence. These topological quantum error-correcting codes
are instances of stabilizer quantum codes \cite{gottesman:96}, in
which errors are diagnosed by measuring certain check operators or
stabilizers. In topological codes these check operators are local,
which, in practice, is an important advantage.  Moreover, error
correction has a deep connection to random spin models in statistical
mechanics and lattice gauge theories \cite{dennis:02-ea}.

\begin{figure}
\includegraphics[width=\columnwidth]{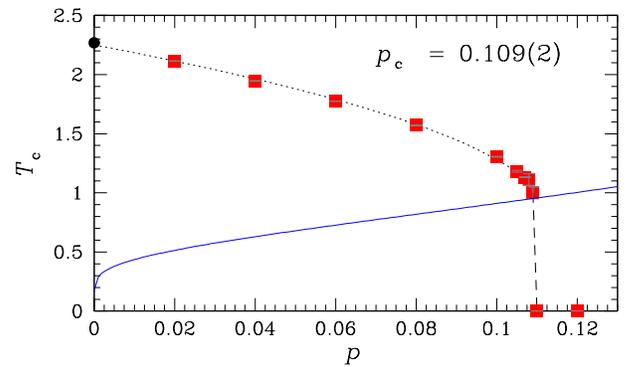}
\vspace*{-1.0cm}
\caption{(Color online)
$p$--$T_c$ phase diagram for the random 3-body Ising model.
For $p > p_c \approx 0.109$ the ferromagnetic order is lost. The
dotted line is a guide to the eye, the black circle represents the
analytically known transition temperature of the 2D Ising model. The
blue (solid) line represents the N-line. In the regime marked
by a dashed line the exact determination of $T_c(p)$ is difficult.
}
\label{fig:pd}
\end{figure}

One of the original motivations for introducing surface codes
was to achieve error protection at the physical level through
energy barriers that would remove the need for external
recovery actions. Only the application of strong magnetic
fields (compared to the topological coupling) destabilizes the
topological phase \cite{trebst:07a-ea}. However, several studies
\cite{dennis:02-ea,alicki:07-ea,alicki:09-ea,alicki:08-ea,iblisdir:09-ea}
and a rigorous proof \cite{alicki:09-ea} have shown that the toric code (TC)
is not stable against thermal excitations, except in four dimensions
\cite{dennis:02-ea,alicki:08-ea}.

Therefore, the study of active error correction in topological
codes \cite{dennis:02-ea} is fully justified.  Ultimately, the
goal is not only to achieve good quantum memories, but also to
perform quantum computations with them. In this regard, the TC
\cite{kitaev:03} is somehow limited since it only allows for a
convenient (transversal) implementation of a limited set of quantum
gates: Pauli gates of $X$ and $Z$ type, and the CNOT gate. To overcome
this limitation, topological color codes (TCC) have been introduced
\cite{bombin:06,bombin:07}. Using TCC, it is possible to implement the
{\em whole} Clifford group of quantum gates and thus realize quantum
distillation, teleportation, etc.  Notice that, although we use the
mapping of Ref.~\cite{dennis:02-ea}, there is a difference regarding
the issue of types of homology involved: our model has a colored
homology, while the Kitaev model has a simple homology. As a result,
unlike the standard Ising model, the resulting statistical mechanical
model has 3-body interactions with a value of $p_c$ a priori unknown,
thus motivating the present study.

The question arises as to whether the wider computational
capabilities of TCCs imply a lower resistance to noise.
We address this problem and show that the (error) threshold
value is $p_c = 0.109(2)$, which is comparable to values for the
TC \cite{honecker:01-ea,merz:02,ohzeki:09}.  To compute $p_c$,
we derive a statistical mechanical model describing the error-correction 
process; a random 3-body Ising model, with (classical)
spins located at the vertices of a triangular lattice. In addition
to thermal fluctuations, the mapping requires the introduction of
quenched randomness to the sign of the interactions that correspond
to faulty bits. One can then study the $p$--$T_c$ phase diagram of
the model, see Fig.~\ref{fig:pd}, where $p$ is the probability for
wrong-sign couplings to appear. For low $T$, $p$ the model orders,
which corresponds to feasible error correction. The critical $p_c$
for error correction is recovered from the critical $p$ along the
Nishimori (N) line \cite{nishimori:81} in the $p$--$T$ plane.

The disordered 3-body Ising model on a triangular lattice has not been
studied before. However, in the absence of randomness it is known to
have a different universality class than the standard Ising model,
but with the same critical temperature \cite{baxter:73}.  Furthermore,
the critical exponents can be computed exactly ($\nu = \alpha = 2/3$),
which allows us to test the numerical results in the $p=0$ limit.

\begin{figure}
\includegraphics[width=\columnwidth]{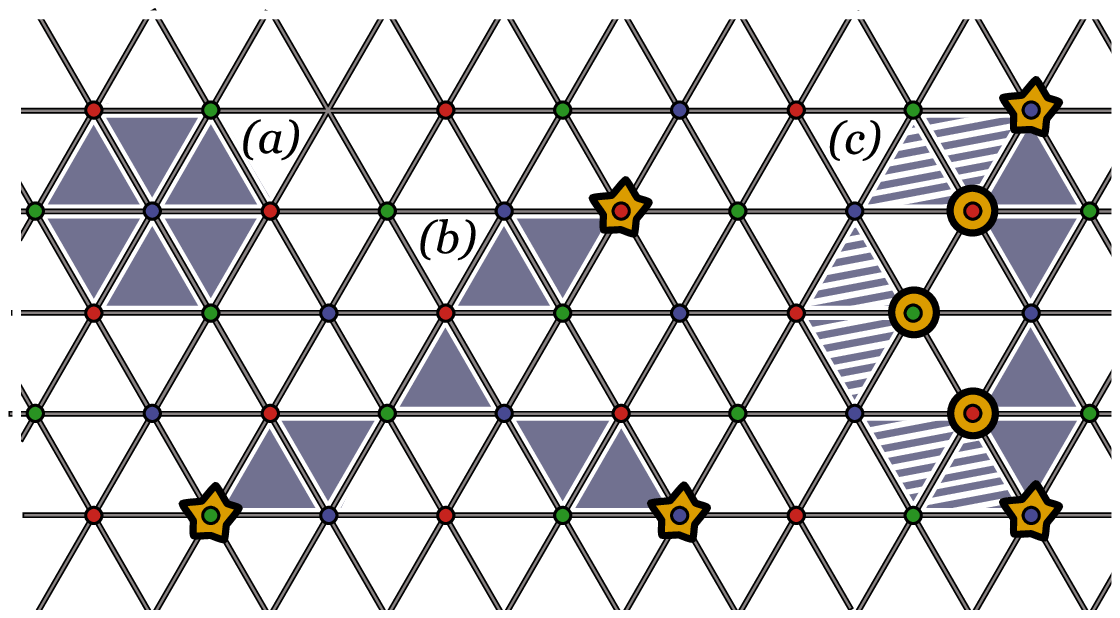}
\vspace*{-0.5cm}
\caption{(Color online)
Lattice for the TCCs with 3-colored vertices. Physical qubits of
the error-correcting code correspond to triangles (stars mark the
`boundaries' of the sets of triangles displayed). (a) Boundary of
a vertex $v$. The stabilizer operators $X_v$ , $Z_v$ have support
on the corresponding qubits. (b) Error pattern in the form of a
string net. The three vertices that form its boundary are all the
information we have to correct the error. (c) Two error patterns
with the same boundary. Because together they form the boundary of
the three vertices marked with a circle, they are equivalent.
}
\label{fig:system}
\end{figure}

\paragraph*{Topological color codes.---}
\label{sec:qec}

To construct a TCC ${\cal C}$ we start from any two-dimensional
(2D) lattice in which all plaquettes are triangles and vertices are
3-colorable, such that no link connects vertices of the same color. The
lattice is embedded in a compact surface of arbitrary topology. Since
information is encoded in topological degrees of freedom, the code
is nontrivial only when the topology of the surface is nontrivial,
e.g., a torus of genus $g \ge 1$.  So far, color codes have been
introduced in the dual lattice (2-colex \cite{bombin:06}). Here we
prefer to work in the triangular lattice to have a more direct mapping,
see Fig.~\ref{fig:system}.

We consider a physical system with a qubit at each lattice triangle, 
and introduce the following vertex operators that generate
the stabilizer group of $\mathcal C$.  For each vertex $v$ we have
two types of operators which correspond to Pauli operators of $X$
or $Z$ type, i.e., $X_v :=\bigotimes_{\triangle : v \in \triangle}
X_{\triangle}$ and $Z_v :=\bigotimes_{\triangle : v \in \triangle}
Z_{\triangle}$.  Thus, a vertex operator acts on all nearby triangles,
see Fig.~\ref{fig:system}. Vertex operators pairwise commute and square
to identity. The code $\mathcal C$ is defined as the subspace with
$X_v=Z_v=1 \; \forall v$. To perform error correction one measures
vertex operators. The resulting collection of $\pm 1$ eigenvalues is
the error syndrome.

\paragraph*{Error correction.---}

Color codes have a structure with stabilizer generators which are
either products of $X$ or $Z$ Pauli operators, but not both. This
allows us to treat bit-flip and phase errors separately, making the
procedure classical: $X$-type ($Z$-type) errors produce violations
of $Z$-type ($X$-type) vertex operators.  Without loss of generality,
let us consider the bit-flip case, that is, errors of the form $X_E:=
\bigotimes_{\triangle \in E} X_\triangle$, where $E$ is the subset
of triangles that suffered a bit-flip. Let $\partial E$ be the
collection of vertices that are part of an odd number of triangles
in $E$, i.e., the boundary of a set of triangles $E$ is chosen so
that the error $X_E$ gives rise to a syndrome with $Z_v=-1$ at those
vertices $v\in \partial E$, see Fig.~\ref{fig:system}. In trying to
correct the error, we apply to the system bit-flips $X_{E^\prime}$
with the same boundary, $\partial E^\prime = \partial E$. This is
only successful as long as $X_{E^\prime}X_E=:X_{E+E^\prime}$ is
an element of the stabilizer group. Geometrically, $D=E+E^\prime$
is a cycle: its boundary $\partial D$ is empty. Given a vertex $v$,
let $\partial v$ be the subset of triangles meeting at $v$. We say
that $D$ is a boundary if $D=\sum_V \partial v$ for some subset of
triangles $V$. In that case, $X_D$ is an element of the stabilizer
group. Thus, error correction is successful whenever $D$ is a boundary,
i.e., if $D$ has trivial homology. In that case the real error $E$
and the guessed error $E^\prime$ belong to the same homology class.

\paragraph*{Mapping to a random 3-body Ising model.---}
\label{sec:mapping}

We consider a standard error model based on stochastic errors in which
phase errors $Z$ and qubit bit-flip errors $X$ are uncorrelated and
occur with probability $p$ at each qubit. We focus on the correction
of bit-flip errors.

Let $P(E)$ be the probability for a given set of bit-flip errors
$E$. Up to a $p$-dependent factor, $P(E)\propto [p/(1-p)]^{|E|}$. We
may also consider the total probability for the corresponding homology
class $\bar E$ of errors, $P(\bar E):=\sum_{D} P(E+D)$, where $D$
runs over all boundaries. If we measure a syndrome $\partial E$,
then the probability that it was caused by an error in the homology
class $\bar E$ is
\begin{equation}
P(\bar E | \partial E) = \frac{P(\bar E)}{\sum_{i} P( {\bar E+\bar
D_i})} ,
\label{class_probability}
\end{equation}
where the $D_i$ are representatives of the homology classes of cycles
\cite{dennis:02-ea}. Then, error correction is achievable if in the
limit of infinite system size we have $\sum_E P(E) \, P(\bar E|\partial
E) \rightarrow 1$.  That is, $p < p_c$ if for those syndromes which
have a nonnegligible probability to appear the error can be guessed
with total confidence.

Following Ref.~\cite{dennis:02-ea}, we set $\exp(-2K):=p/(1-p)$ (with
$K = J/T$, $T$ the temperature) for the N-line so that $P(E)
\propto \exp(K \sum_\triangle \tau_\triangle)$, where the sum is
over all the triangular plaquettes (qubits) and $\tau_\triangle=\pm
1 < 0$ when $\triangle\in E$. By inserting classical spin
variables $\sigma_i = \pm 1$ at the vertices and labeling the triangles
$\triangle$ with triplets of vertices $\langle i j k\rangle$ we write
$P(\bar E)$ as a partition function
\begin{equation}
P(\bar E)\propto Z[K,\vect \tau]:=\sum_{\vect \sigma}  e^{K \sum_{\langle i j k\rangle} \tau_{ijk} \sigma_i\sigma_j\sigma_k}.
\label{partition_function}
\end{equation}
Equation (\ref{partition_function}) is a 3-body classical Ising
model with the couplings' sign given by $\vect\tau$. When all 
$\tau_\triangle = 1$ the model is ferromagnetically ordered at
low $T$. Negative $\tau_\triangle$ introduce frustration in the form
of {\em nets} of domain walls. These can branch, a new feature not
present in the random bond Ising model associated with the TC.

The relative importance of the different error homology classes $P(\bar
E+\bar D_i)=Z[K,\vect \tau_i]$ in \eqref{class_probability} is given
by the free energy cost of introducing a domain wall $D_i$, because
\begin{equation}
\Delta_i(\vect \tau)=\beta F(K,\vect \tau_i)-\beta F(K,\vect \tau)=\ln \left ( \frac {Z[K,\vect \tau]}{Z[K,\vect \tau_i]} \right ).
\label{wall_cost}
\end{equation}
The cost $\Delta_i$ must be averaged over all coupling configurations,
with $p$ the probability for any triangle to have $\tau_\triangle=
-1$. Thus we are led to the study of a random 3-body Ising model.
For low $p$ and $T$ (high $K = J/T$) the system is ordered and
domain-wall fluctuations are suppressed: $\Delta_i$ diverges with
the system size for nontrivial domain walls. The critical error
threshold $p_c$ for error correction is recovered from the $p$--$T$
phase diagram as the critical $p$ along the N-line $e^{-2J/T}=p/(1-p)$
\cite{comment:J}.

\begin{figure*}[!tb]
\includegraphics[width=0.5\columnwidth]{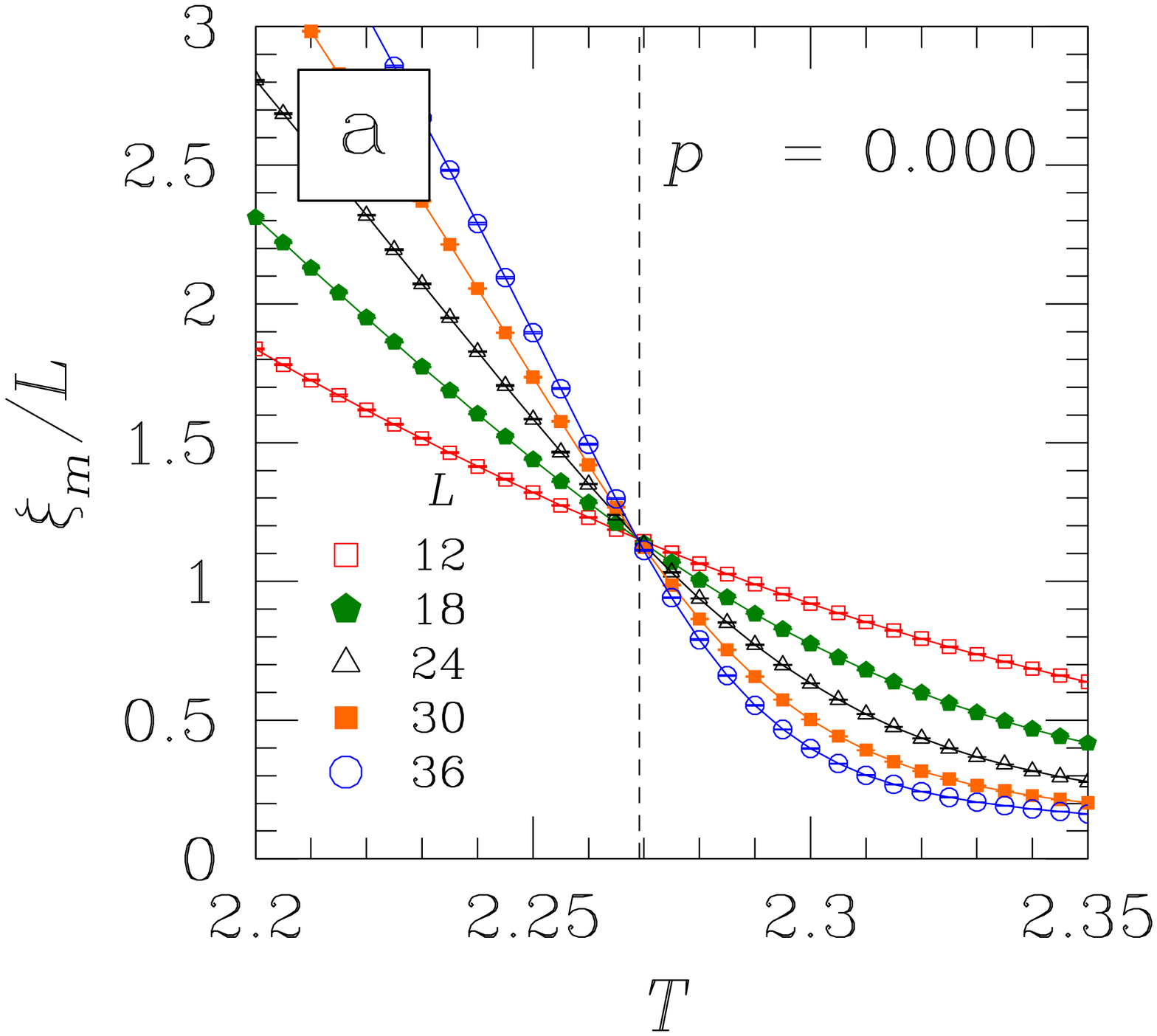}
\hspace*{-0.45cm}
\includegraphics[width=0.5\columnwidth]{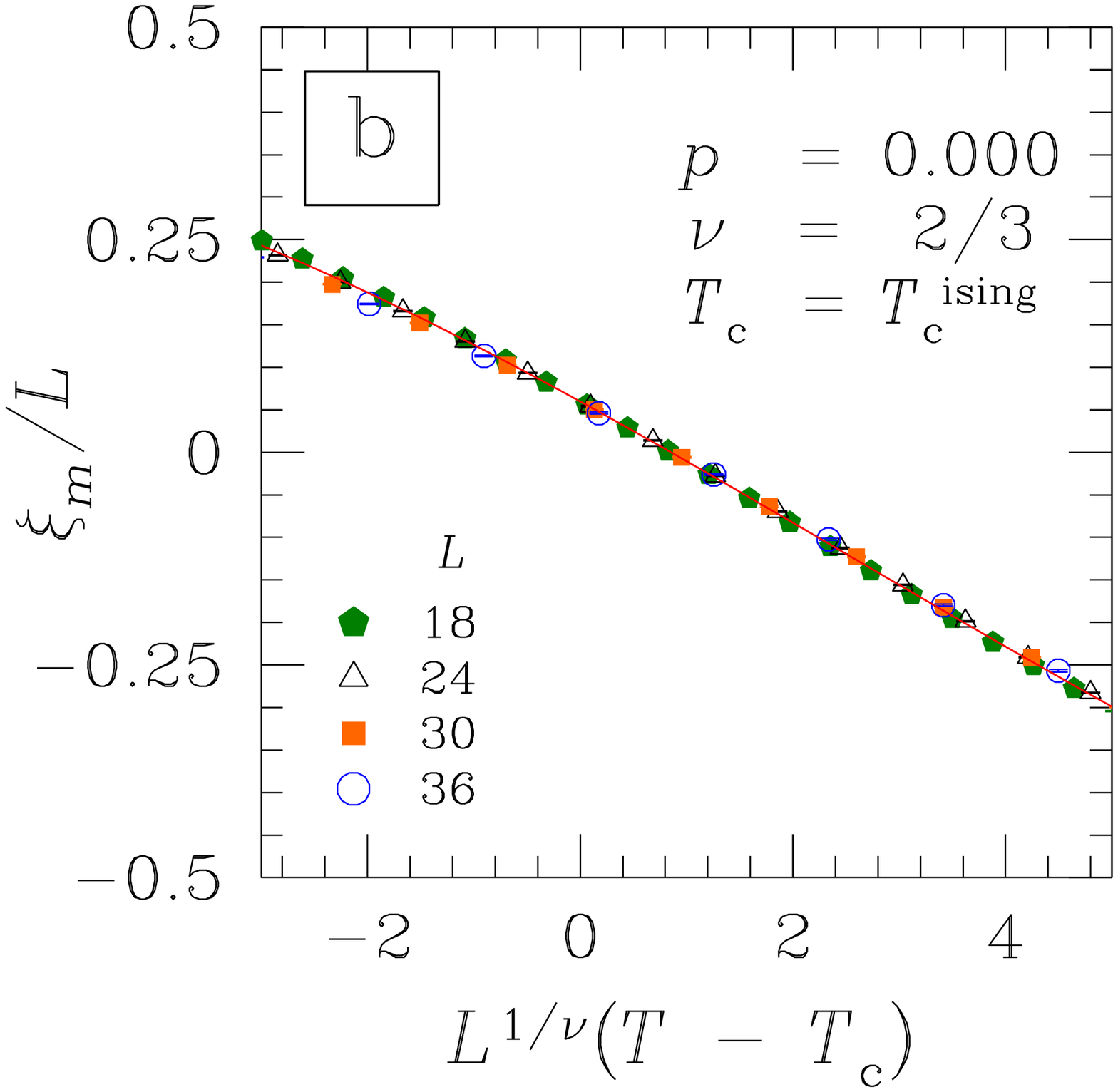}
\hspace*{-0.45cm}
\includegraphics[width=0.5\columnwidth]{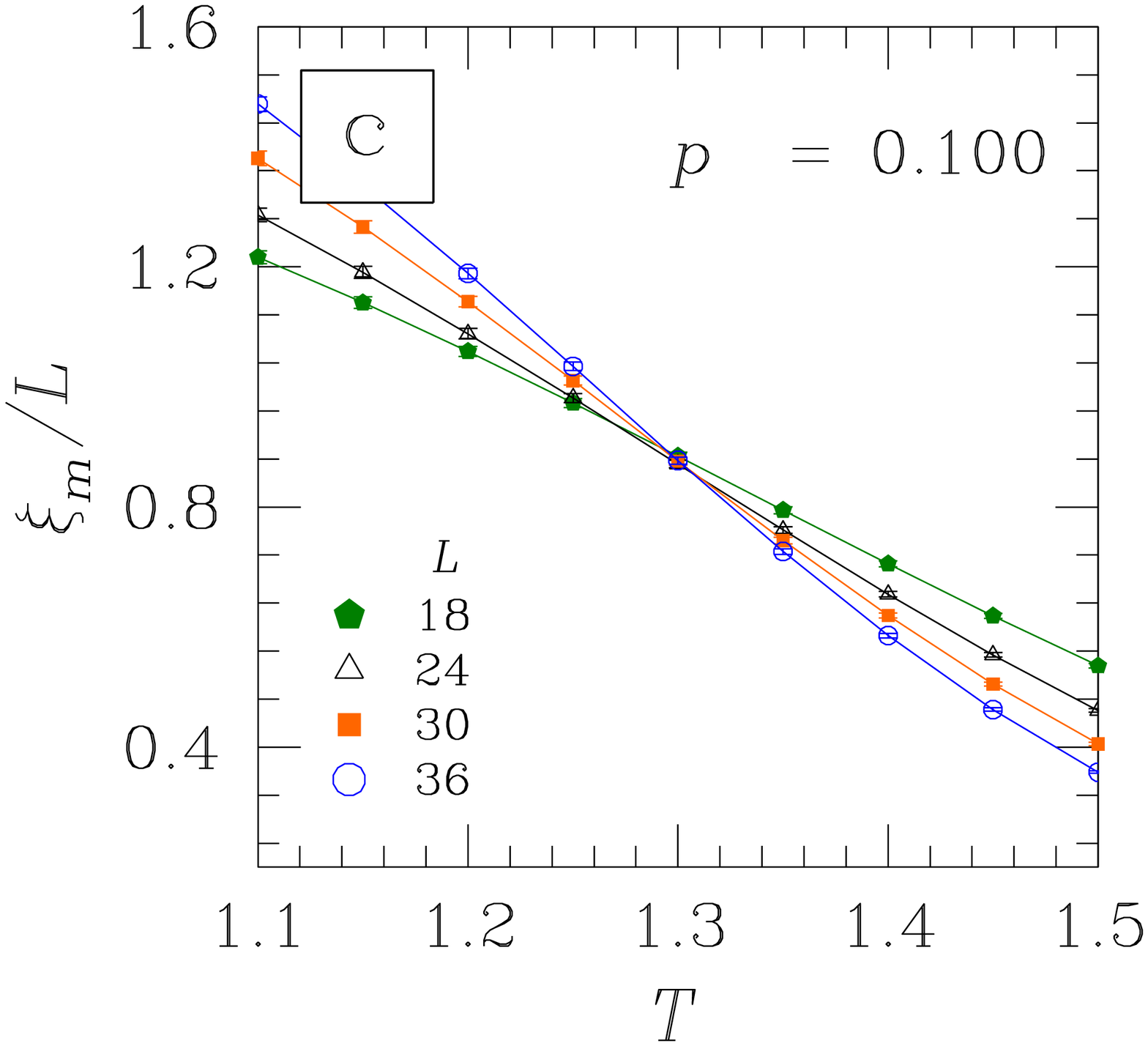}
\hspace*{-0.45cm}
\includegraphics[width=0.5\columnwidth]{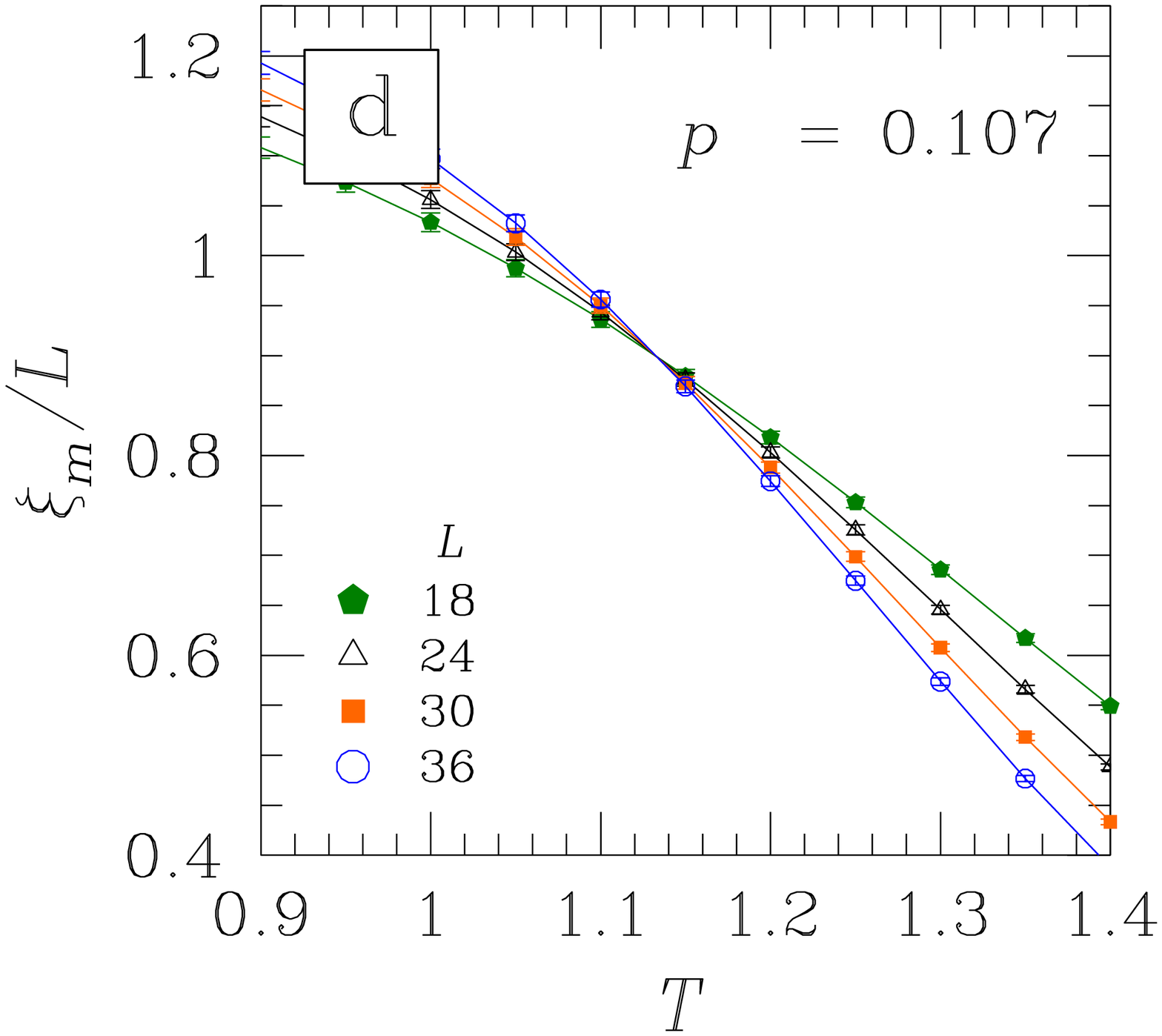}
\vspace*{-0.45cm}
\includegraphics[width=0.5\columnwidth]{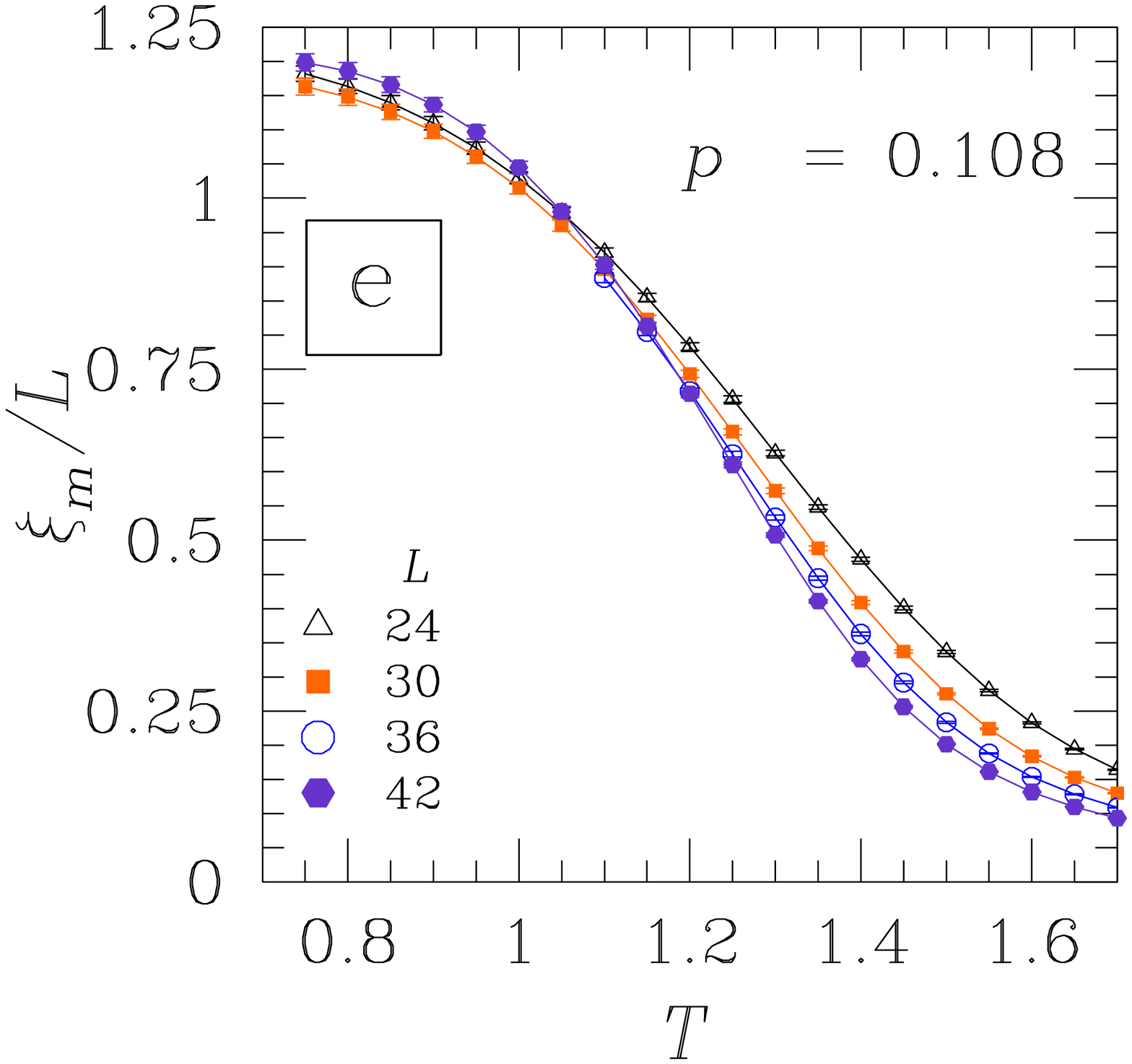}
\hspace*{-0.45cm}
\includegraphics[width=0.5\columnwidth]{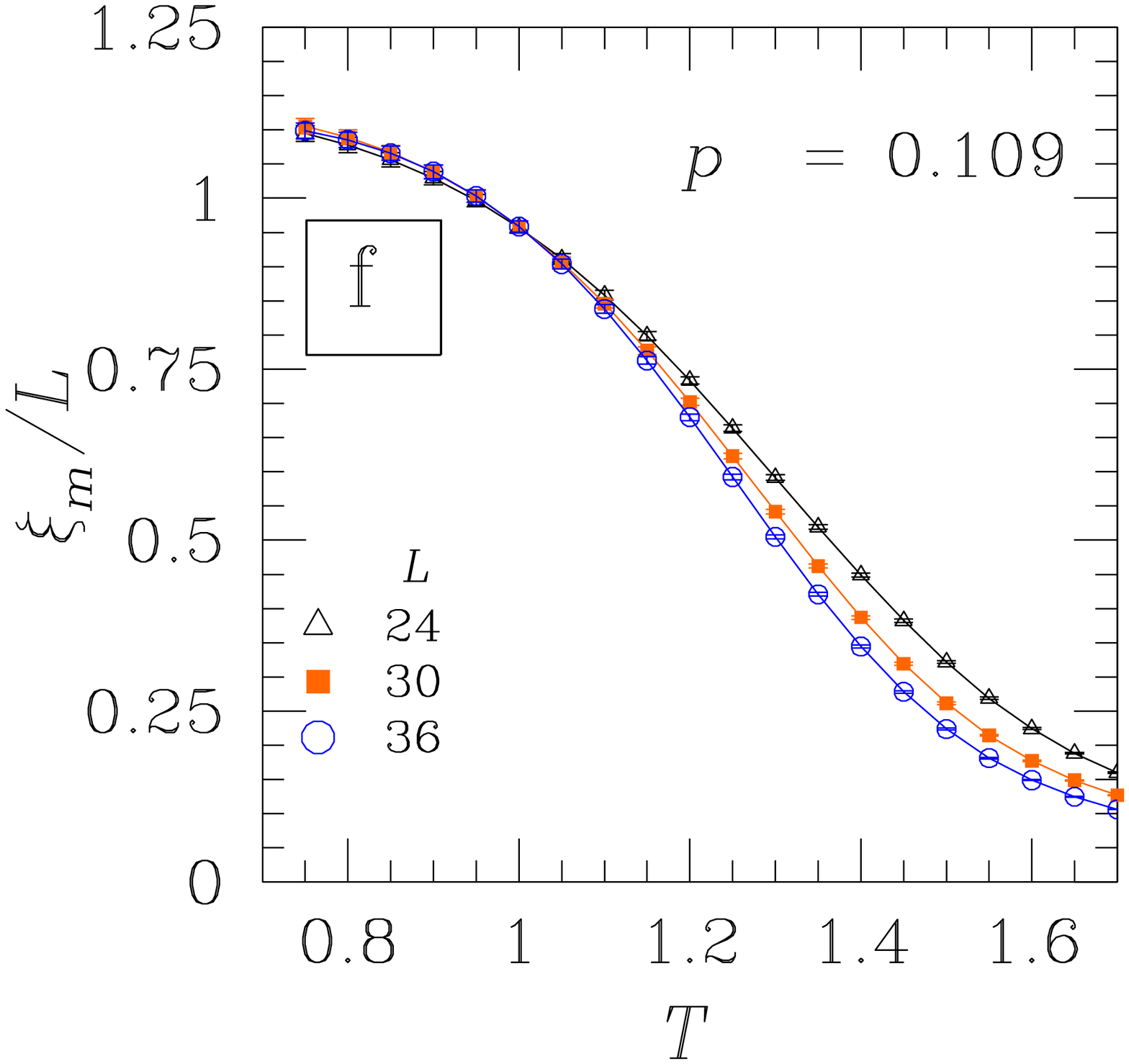}
\hspace*{-0.45cm}
\includegraphics[width=0.5\columnwidth]{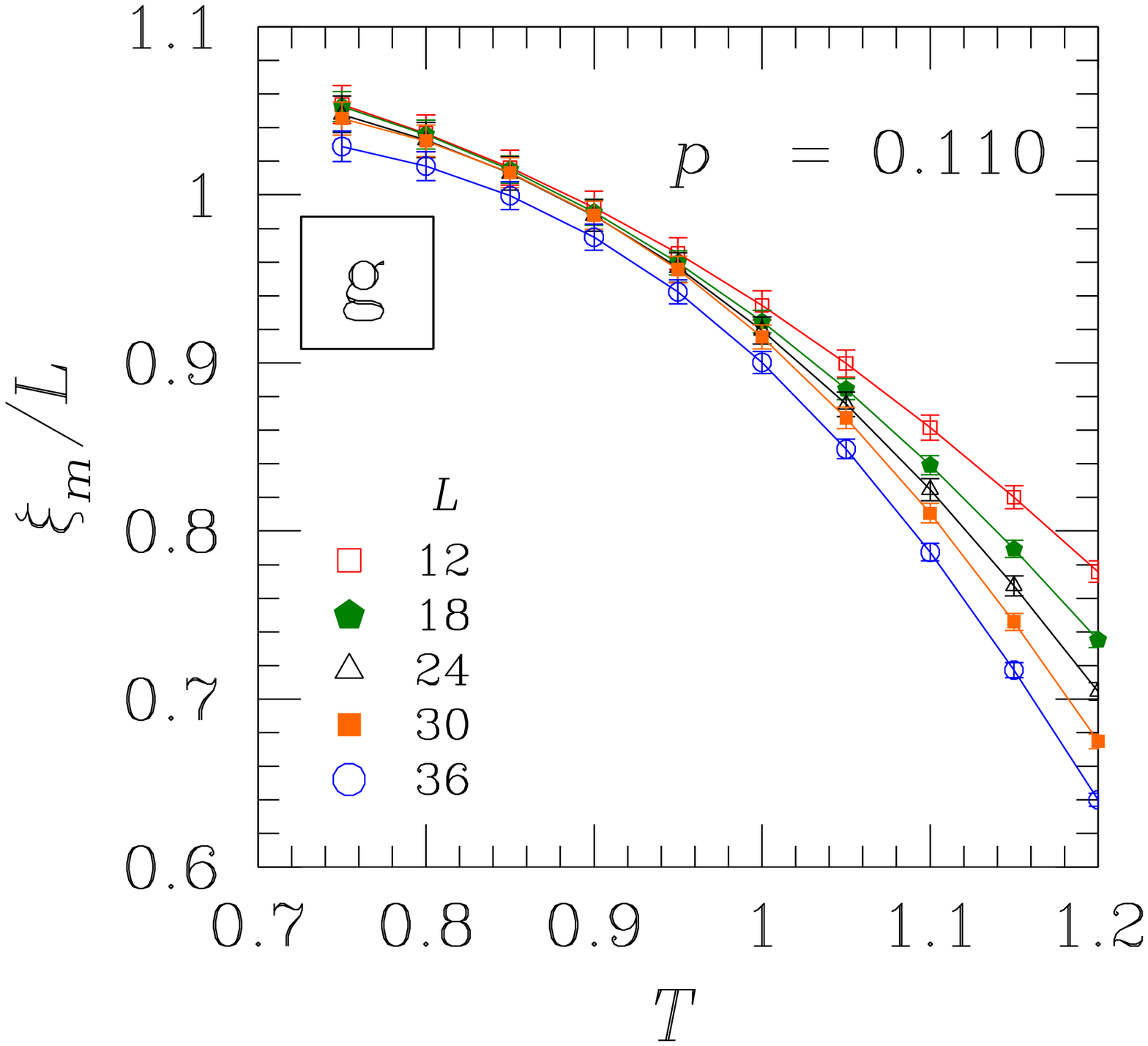}
\hspace*{-0.45cm}
\includegraphics[width=0.5\columnwidth]{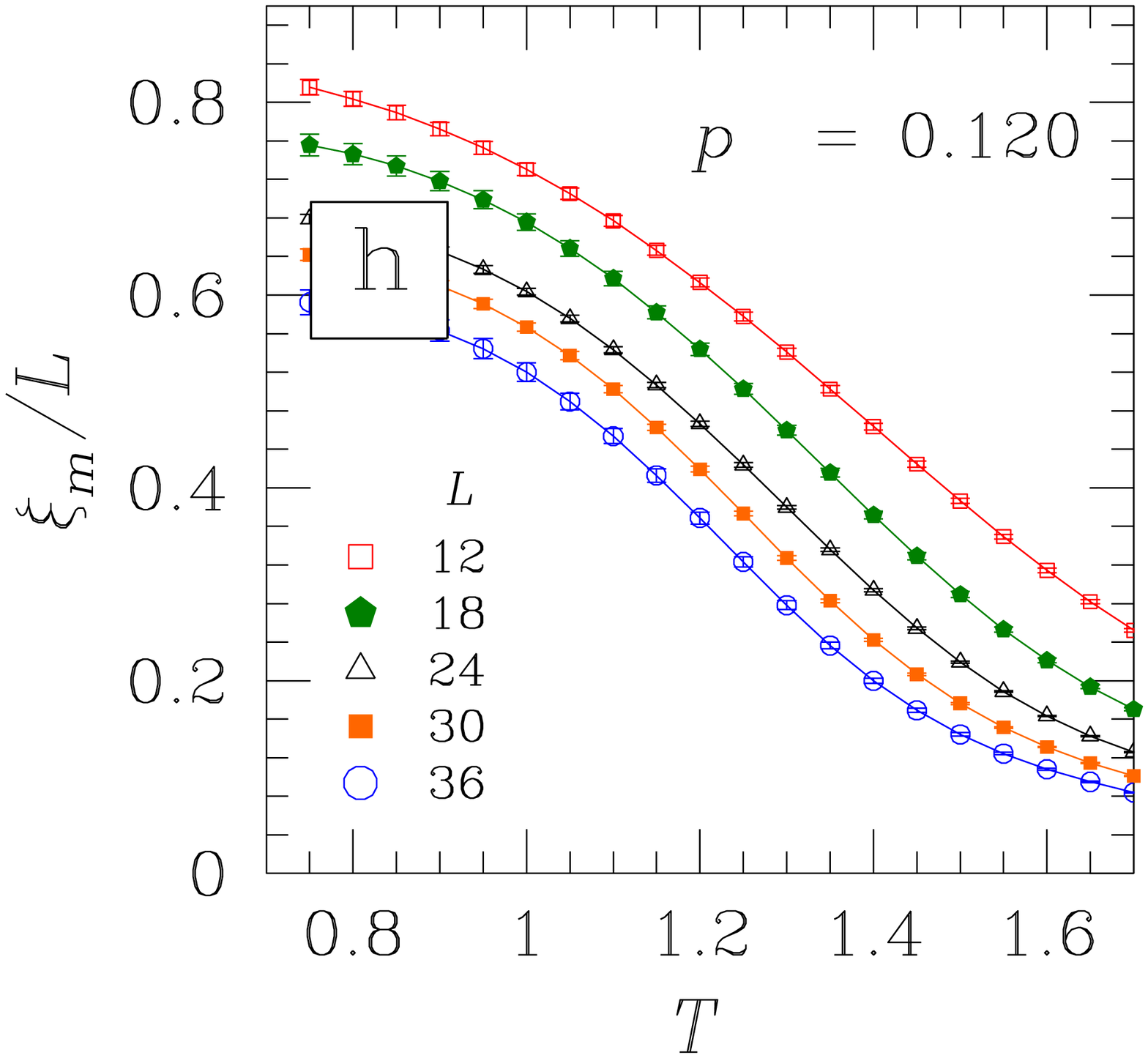}

\caption{(Color online)
Finite-size correlation length $\xi_{\rm m}/L$ as a function of
temperature $T$ for different values of $p$. (a) $p = 0$. The data
cross at the critical temperature of the 2D Ising model (dashed
line). (b) Finite-size scaling analysis of the data for $p = 0$
using $\nu = 2/3$. The scaling is very good showing that corrections
to scaling are negligible. (c)---(f) For $p \lesssim p_c = 0.109$
there is signature of a transition (data for different $L$ cross)
whereas for $p > p_c$ the transition vanishes [panels (g)---(h)].
}
\label{fig:xi}
\end{figure*}

\paragraph*{Numerical details.---}
\label{sec:numerics}

\begin{table}[!tb]
\caption{
Simulation parameters: $L$ is the system size, $N_{\rm sa}$ is the
number of disorder samples, $t_{\rm eq} = 2^{b}$ is the number of
equilibration sweeps, $T_{\rm min}$ [$T_{\rm max}$] is the lowest
[highest] temperature, and $N_{\rm T}$ the number of temperatures used.
\label{tab:simparams}}
{\footnotesize
\begin{tabular*}{\columnwidth}{@{\extracolsep{\fill}} l r r r r r r}
\hline
\hline
$p$ & $L$ & $N_{\rm sa}$ & $b$ & $T_{\rm min}$ & $T_{\rm max}$ &$N_{\rm T}$ \\
\hline
$0.00$            & $12$, $18$ & $20$     &  $18$ & $2.200$ & $2.350$ & $31$ \\
$0.00$            & $24$, $30$ & $20$     &  $19$ & $2.200$ & $2.350$ & $31$ \\
$0.00$            &       $36$ & $20$     &  $20$ & $2.200$ & $2.350$ & $31$ \\
$0.02$            & $12$, $18$ & $5\,000$ &  $18$ & $1.900$ & $2.400$ & $51$ \\
$0.02$            & $24$, $30$ & $5\,000$ &  $19$ & $1.900$ & $2.400$ & $51$ \\
$0.02$            &       $36$ & $5\,000$ &  $20$ & $1.900$ & $2.400$ & $51$ \\
$0.04$            & $12$, $18$ & $5\,000$ &  $18$ & $1.700$ & $2.200$ & $51$ \\
$0.04$            & $24$, $30$ & $5\,000$ &  $19$ & $1.700$ & $2.200$ & $51$ \\
$0.04$            &       $36$ & $5\,000$ &  $20$ & $1.700$ & $2.200$ & $51$ \\
$0.06$            & $12$, $18$ & $5\,000$ &  $18$ & $1.600$ & $2.100$ & $51$ \\
$0.06$            & $24$, $30$ & $5\,000$ &  $19$ & $1.600$ & $2.100$ & $51$ \\
$0.06$            &       $36$ & $5\,000$ &  $20$ & $1.600$ & $2.100$ & $51$ \\
$0.08$            & $12$, $18$ & $5\,000$ &  $18$ & $1.400$ & $2.000$ & $61$ \\
$0.08$            & $24$, $30$ & $5\,000$ &  $19$ & $1.400$ & $2.000$ & $61$ \\
$0.08$            &       $36$ & $5\,000$ &  $20$ & $1.400$ & $2.000$ & $61$ \\
$0.10$ --- $0.12$ & $12$, $18$ & $5\,000$ &  $18$ & $0.750$ & $2.600$ & $38$ \\
$0.10$ --- $0.12$ & $24$, $30$ & $5\,000$ &  $19$ & $0.750$ & $2.600$ & $38$ \\
$0.10$ --- $0.12$ &       $36$ & $5\,000$ &  $20$ & $0.750$ & $2.600$ & $38$ \\
\hline
\hline
\end{tabular*}
}
\end{table}

To determine the existence of a ferromagnetic phase we compute the
finite-size correlation length \cite{palassini:99b}.  We start by
determining the wave-vector-dependent susceptibility given by $\chi(k)
=  (1/L^2) \sum_{ij} \langle S_iS_j \rangle_T \exp[i{\bf k}\cdot({\bf
R}_i - {\bf R}_j)]$.  Here $\langle \cdots \rangle_T$ denotes a
thermal average and ${\bf R}_i$ the spatial location of the spins.
The correlation length is given by
\begin{equation}
\xi_{\rm m} = (1/2)\sin^{-1}(q/2)
\sqrt{[\chi(0)]_{\rm av}/[\chi(q)]_{\rm av}
- 1} ,
\label{eq:xiL}
\end{equation}
where $q = (2 \pi / L,0)$ is the smallest nonzero wave-vector and
$[\cdots]_{\rm av}$ represents an average over $N_{\rm sa}$ disorder
(error) samples. $\xi_{\rm m}/L \sim \widetilde{X}(L^{1/\nu}[T -
T_{\rm c}] )$, i.e., if there is a transition at $T = T_c$, data for
${\xi_{\rm m}}/{L}$ for different system sizes $L$ cross at $T_c$
[see, for example Fig.~\ref{fig:xi}(a)].  The critical exponent $\nu$
for the correlation length can be determined by a full scaling of the
data, as shown in Fig.~\ref{fig:xi}(b).  We also probe the existence of
a spin-glass phase by computing the spin-glass finite-size correlation
length.

The disorder in Eq.~\eqref{partition_function} increases the numerical
complexity of the problem drastically with a behavior reminiscent of
spin glasses \cite{binder:86}.  To speed up the simulations, we use
the exchange Monte Carlo method \cite{hukushima:96}. Equilibration
is tested by a logarithmic binning of the data. Once the last three
bins agree within errors, we define the system to be equilibrated.
Simulation parameters are shown in Table \ref{tab:simparams}.

\paragraph*{Error threshold.---}
\label{sec:thresh}

Figure \ref{fig:xi} shows the temperature-dependent finite-size
correlation length for different values of $p$. (a) Data for
$p = 0$, the ferromagnetic case. The dashed line represents the
transition temperature of the 2D Ising model $T_c \simeq 2.2692$
\cite{yeomans:92}.  The agreement with the numerical data is
excellent, suggesting that corrections to scaling are negligible.
(b) Finite-size scaling analysis of the data in (a) using the exact
exponent $\nu = 2/3$. (c) --- (h) Finite-size correlation length
for different $p$ values. For $p = 0.108$ marginal behavior appears
and the determination of the transition is difficult.  Because $p =
0.107$ shows a transition, and $p = 0.109$ shows marginal behavior,
whereas $p = 0.110$ shows no sign of a transition, we conservatively
estimate $p_c = 0.109(2)$ \cite{comment:T0}. This is close to
estimates for the TC where $p_c^{\rm TC}$ has been continuously
improved from $0.1094(2)$ \cite{honecker:01-ea} to $0.1093(2)$
\cite{merz:02} and $0.109187$ \cite{ohzeki:09}.  The $p$--$T_c$
phase diagram is shown in Fig.~\ref{fig:pd}; the solid (blue)
line being the N-line. We have also verified that there is no
spin-glass order in the model (not shown).  Finally, we ensure
that our results do not violate the quantum Gilbert-Varshamov
bound \cite{honecker:01-ea,merz:02,ohzeki:09,comment:css} where
the encoding rate $R(p)$ must satisfy $R(p) \leq 1 - 2 H(p)$,
$H(p) = -p \log_2(p) - (1-p) \log_2(1-p)$ the Shannon entropy,
\cite{gilbert:52,varshamov:57,calderbank:96}. For our estimate the
bound is satisfied, since it lies under the zero-rate probability $p
\simeq 0.110027$.

\paragraph*{Conclusions.---}
\label{sec:conlcusions}

In summary, we have computed the error threshold for TCCs
on a triangular lattice by mapping the problem onto a 3-body
random Ising model on a triangular lattice. Using Monte Carlo
simulations we find for the error threshold $p_c = 0.109(2)$
\cite{comment:other}. Therefore, TCCs are as robust as the Kitaev
toric code with the added benefit of being able to represent the
whole Clifford group of quantum gates. The studied 3-body random
Ising model highlights the relationship between spin-glass physics and
information theory \cite{nishimori:01}, e.g., fully connected systems,
and presents a new class of system exhibiting glassy behavior via
3-body interactions, without spin-reversal symmetry.  Future work
will focus on the impact of faulty measurements and the corresponding
mapping to a $(2+1)$-dimensional random gauge model.

\begin{acknowledgments} 

We thank A.~F.~Albuquerque and A.~Landahl for useful discussions.
M.A.M.-D.~and H.B.~acknowledge financial support from a PFI grant
of EJ-GV, DGS grants under contracts, FIS2006-04885, and the ESF
INSTANS 2005-10.  H.G.K.~acknowledges support from the SNF (Grant
No.~PP002-114713).  The authors acknowledge the Texas Advanced
Computing Center (TACC) at The University of Texas at Austin for
providing HPC resources (Ranger Sun Constellation Linux Cluster), the
Centro de Supercomputaci{\'o}n y Visualizaci{\'o}n de Madrid (CeSViMa)
for access to the magerit cluster, the Barcelona Supercomputing Center
for access to the MareNostrum cluster within the Spanish Supercomputing
Network and ETH Zurich for CPU time on the Brutus cluster.

\end{acknowledgments}

\bibliography{refs,comments}

\begin{thebibliography}{32}
\expandafter\ifx\csname natexlab\endcsname\relax\def\natexlab#1{#1}\fi
\expandafter\ifx\csname bibnamefont\endcsname\relax
  \def\bibnamefont#1{#1}\fi
\expandafter\ifx\csname bibfnamefont\endcsname\relax
  \def\bibfnamefont#1{#1}\fi
\expandafter\ifx\csname citenamefont\endcsname\relax
  \def\citenamefont#1{#1}\fi
\expandafter\ifx\csname url\endcsname\relax
  \def\url#1{\texttt{#1}}\fi
\expandafter\ifx\csname urlprefix\endcsname\relax\def\urlprefix{URL }\fi
\providecommand{\bibinfo}[2]{#2}
\providecommand{\eprint}[2][]{\url{#2}}

\bibitem[{\citenamefont{Shor}(1995)}]{shor:95}
\bibinfo{author}{\bibfnamefont{P.~W.} \bibnamefont{Shor}},
  \bibinfo{journal}{Phys. Rev. A} \textbf{\bibinfo{volume}{52}},
  \bibinfo{pages}{R2493} (\bibinfo{year}{1995}).

\bibitem[{\citenamefont{Steane}(1996)}]{steane:96}
\bibinfo{author}{\bibfnamefont{A.~M.} \bibnamefont{Steane}},
  \bibinfo{journal}{Phys. Rev. Lett.} \textbf{\bibinfo{volume}{77}},
  \bibinfo{pages}{793} (\bibinfo{year}{1996}).

\bibitem[{\citenamefont{Kitaev}(2003)}]{kitaev:03}
\bibinfo{author}{\bibfnamefont{A.~Y.} \bibnamefont{Kitaev}},
  \bibinfo{journal}{Ann. Phys.} \textbf{\bibinfo{volume}{303}},
  \bibinfo{pages}{2} (\bibinfo{year}{2003}).

\bibitem[{\citenamefont{Gottesman}(1996)}]{gottesman:96}
\bibinfo{author}{\bibfnamefont{D.}~\bibnamefont{Gottesman}},
  \bibinfo{journal}{Phys. Rev. A} \textbf{\bibinfo{volume}{54}},
  \bibinfo{pages}{1862} (\bibinfo{year}{1996}).

\bibitem[{\citenamefont{Dennis~{\em et al.}}(2002)}]{dennis:02-ea}
\bibinfo{author}{\bibfnamefont{E.}~\bibnamefont{Dennis~{\em et al.}}},
  \bibinfo{journal}{J. Math. Phys.} \textbf{\bibinfo{volume}{43}},
  \bibinfo{pages}{4452} (\bibinfo{year}{2002}).

\bibitem[{\citenamefont{Trebst~{\em et al.}}(2007)}]{trebst:07a-ea}
\bibinfo{author}{\bibfnamefont{S.}~\bibnamefont{Trebst~{\em et al.}}},
  \bibinfo{journal}{Phys. Rev. Lett} \textbf{\bibinfo{volume}{98}},
  \bibinfo{pages}{070602} (\bibinfo{year}{2007}).

\bibitem[{\citenamefont{Alicki~{\em et al.}}(2007)}]{alicki:07-ea}
\bibinfo{author}{\bibfnamefont{R.}~\bibnamefont{Alicki~{\em et al.}}},
  \bibinfo{journal}{J. Phys. A} \textbf{\bibinfo{volume}{40}},
  \bibinfo{pages}{6451} (\bibinfo{year}{2007}).

\bibitem[{\citenamefont{Alicki~{\em et al.}}(2009)}]{alicki:09-ea}
\bibinfo{author}{\bibfnamefont{R.}~\bibnamefont{Alicki~{\em et al.}}},
  \bibinfo{journal}{J. Phys. A} \textbf{\bibinfo{volume}{42}},
  \bibinfo{pages}{065303} (\bibinfo{year}{2009}).

\bibitem[{\citenamefont{Alicki~{\em et al.}}(2008)}]{alicki:08-ea}
\bibinfo{author}{\bibfnamefont{R.}~\bibnamefont{Alicki~{\em et al.}}}
  (\bibinfo{year}{2008}), \bibinfo{note}{(arXiv:quant-phys/0811.0033)}.

\bibitem[{\citenamefont{Iblisdir~{\em et al.}}(2009)}]{iblisdir:09-ea}
\bibinfo{author}{\bibfnamefont{S.}~\bibnamefont{Iblisdir~{\em et al.}}},
  \bibinfo{journal}{Phys. Rev. B} \textbf{\bibinfo{volume}{79}},
  \bibinfo{pages}{134303} (\bibinfo{year}{2009}).

\bibitem[{\citenamefont{Bombin and Martin-Delgado}(2006)}]{bombin:06}
\bibinfo{author}{\bibfnamefont{H.}~\bibnamefont{Bombin}} \bibnamefont{and}
  \bibinfo{author}{\bibfnamefont{M.~A.} \bibnamefont{Martin-Delgado}},
  \bibinfo{journal}{Phys. Rev. Lett.} \textbf{\bibinfo{volume}{97}},
  \bibinfo{pages}{180501} (\bibinfo{year}{2006}).

\bibitem[{\citenamefont{Bombin and Martin-Delgado}(2007)}]{bombin:07}
\bibinfo{author}{\bibfnamefont{H.}~\bibnamefont{Bombin}} \bibnamefont{and}
  \bibinfo{author}{\bibfnamefont{M.~A.} \bibnamefont{Martin-Delgado}},
  \bibinfo{journal}{Phys. Rev. B} \textbf{\bibinfo{volume}{75}},
  \bibinfo{pages}{075103} (\bibinfo{year}{2007}).

\bibitem[{\citenamefont{Honecker~{\em et al.}}(2001)}]{honecker:01-ea}
\bibinfo{author}{\bibfnamefont{A.}~\bibnamefont{Honecker~{\em et al.}}},
  \bibinfo{journal}{Phys. Rev. Lett.} \textbf{\bibinfo{volume}{87}},
  \bibinfo{pages}{047201} (\bibinfo{year}{2001}).

\bibitem[{\citenamefont{Merz and Chalker}(2002)}]{merz:02}
\bibinfo{author}{\bibfnamefont{F.}~\bibnamefont{Merz}} \bibnamefont{and}
  \bibinfo{author}{\bibfnamefont{J.~T.} \bibnamefont{Chalker}},
  \bibinfo{journal}{Phys. Rev. B} \textbf{\bibinfo{volume}{65}},
  \bibinfo{pages}{054425} (\bibinfo{year}{2002}).

\bibitem[{\citenamefont{Ohzeki}(2009{\natexlab{a}})}]{ohzeki:09}
\bibinfo{author}{\bibfnamefont{M.}~\bibnamefont{Ohzeki}},
  \bibinfo{journal}{Phys. Rev. E} \textbf{\bibinfo{volume}{79}},
  \bibinfo{pages}{021129} (\bibinfo{year}{2009}{\natexlab{a}}).

\bibitem[{\citenamefont{{Nishimori}}(1981)}]{nishimori:81}
\bibinfo{author}{\bibfnamefont{H.}~\bibnamefont{{Nishimori}}},
  \bibinfo{journal}{Prog. Theor. Phys.} \textbf{\bibinfo{volume}{66}},
  \bibinfo{pages}{1169} (\bibinfo{year}{1981}).

\bibitem[{\citenamefont{{Baxter} and {Wu}}(1973)}]{baxter:73}
\bibinfo{author}{\bibfnamefont{R.~J.} \bibnamefont{{Baxter}}} \bibnamefont{and}
  \bibinfo{author}{\bibfnamefont{F.~Y.} \bibnamefont{{Wu}}},
  \bibinfo{journal}{Phys. Rev. Lett.} \textbf{\bibinfo{volume}{31}},
  \bibinfo{pages}{1294} (\bibinfo{year}{1973}).

\bibitem[{com({\natexlab{a}})}]{comment:J}
\bibinfo{note}{Without loss of generality, we set the energy scale $J = 1$.}

\bibitem[{\citenamefont{Palassini and Caracciolo}(1999)}]{palassini:99b}
\bibinfo{author}{\bibfnamefont{M.}~\bibnamefont{Palassini}} \bibnamefont{and}
  \bibinfo{author}{\bibfnamefont{S.}~\bibnamefont{Caracciolo}},
  \bibinfo{journal}{Phys. Rev. Lett.} \textbf{\bibinfo{volume}{82}},
  \bibinfo{pages}{5128} (\bibinfo{year}{1999}).

\bibitem[{\citenamefont{Binder and Young}(1986)}]{binder:86}
\bibinfo{author}{\bibfnamefont{K.}~\bibnamefont{Binder}} \bibnamefont{and}
  \bibinfo{author}{\bibfnamefont{A.~P.} \bibnamefont{Young}},
  \bibinfo{journal}{Rev. Mod. Phys.} \textbf{\bibinfo{volume}{58}},
  \bibinfo{pages}{801} (\bibinfo{year}{1986}).

\bibitem[{\citenamefont{Hukushima and Nemoto}(1996)}]{hukushima:96}
\bibinfo{author}{\bibfnamefont{K.}~\bibnamefont{Hukushima}} \bibnamefont{and}
  \bibinfo{author}{\bibfnamefont{K.}~\bibnamefont{Nemoto}},
  \bibinfo{journal}{J. Phys. Soc. Jpn.} \textbf{\bibinfo{volume}{65}},
  \bibinfo{pages}{1604} (\bibinfo{year}{1996}).

\bibitem[{\citenamefont{Yeomans}(1992)}]{yeomans:92}
\bibinfo{author}{\bibfnamefont{J.~M.} \bibnamefont{Yeomans}},
  \emph{\bibinfo{title}{{Statistical Mechanics of Phase Transitions}}}
  (\bibinfo{publisher}{Oxford University Press}, \bibinfo{address}{Oxford},
  \bibinfo{year}{1992}).

\bibitem[{com({\natexlab{b}})}]{comment:T0}
\bibinfo{note}{Computing the phase boundary at $T = 0$ to determine a putative
  reentrant behavior requires different numerical methods. This will be done in
  a subsequent study.}

\bibitem[{com({\natexlab{c}})}]{comment:css}
\bibinfo{note}{The Gilbert-Varshamov bound also works for TCCs since they are
  quantum Calderbank-Shor-Steane codes \cite{nishimori:04}.}

\bibitem[{\citenamefont{Gilbert}(1952)}]{gilbert:52}
\bibinfo{author}{\bibfnamefont{E.~N.} \bibnamefont{Gilbert}},
  \bibinfo{journal}{Bell System Tech. J.} \textbf{\bibinfo{volume}{31}},
  \bibinfo{pages}{504} (\bibinfo{year}{1952}).

\bibitem[{\citenamefont{Varshamov}(1957)}]{varshamov:57}
\bibinfo{author}{\bibfnamefont{R.~R.} \bibnamefont{Varshamov}},
  \bibinfo{journal}{Dokl. Akad. Nauk} \textbf{\bibinfo{volume}{117}},
  \bibinfo{pages}{739} (\bibinfo{year}{1957}).

\bibitem[{\citenamefont{Calderbank and Shor}(1996)}]{calderbank:96}
\bibinfo{author}{\bibfnamefont{A.~R.} \bibnamefont{Calderbank}}
  \bibnamefont{and} \bibinfo{author}{\bibfnamefont{P.~W.} \bibnamefont{Shor}},
  \bibinfo{journal}{Phys. Rev. A} \textbf{\bibinfo{volume}{54}},
  \bibinfo{pages}{1098} (\bibinfo{year}{1996}).

\bibitem[{com({\natexlab{d}})}]{comment:other}
\bibinfo{note}{Upon completion of this work we were informed that a computation
  of the error threshold for TCCs on a union-jack lattice using $T=0$ methods
  \cite{landahl:09} yielding a lower bound for $p_c$ agrees with our Monte
  Carlo data. Similar results were obtained by a subsequent study using an
  approximate duality argument \cite{ohzeki:09a}}.

\bibitem[{\citenamefont{Nishimori}(2001)}]{nishimori:01}
\bibinfo{author}{\bibfnamefont{H.}~\bibnamefont{Nishimori}},
  \emph{\bibinfo{title}{{Statistical Physics of Spin Glasses and Information
  Processing: An Introduction}}} (\bibinfo{publisher}{Oxford University Press},
  \bibinfo{address}{New York}, \bibinfo{year}{2001}).

\bibitem[{\citenamefont{Nishimori and Sollich}(2004)}]{nishimori:04}
\bibinfo{author}{\bibfnamefont{H.}~\bibnamefont{Nishimori}} \bibnamefont{and}
  \bibinfo{author}{\bibfnamefont{P.}~\bibnamefont{Sollich}},
  \bibinfo{journal}{J. Phys. Soc. Jpn.} \textbf{\bibinfo{volume}{73}},
  \bibinfo{pages}{2701} (\bibinfo{year}{2004}).

\bibitem[{\citenamefont{Landahl et~al.}()\citenamefont{Landahl, Anderson, and
  Rice}}]{landahl:09}
\bibinfo{author}{\bibfnamefont{A.}~\bibnamefont{Landahl}},
  \bibinfo{author}{\bibfnamefont{J.~T.} \bibnamefont{Anderson}},
  \bibnamefont{and} \bibinfo{author}{\bibfnamefont{P.}~\bibnamefont{Rice}},
  \bibinfo{note}{in preparation (2009)}.

\bibitem[{\citenamefont{Ohzeki}(2009{\natexlab{b}})}]{ohzeki:09a}
\bibinfo{author}{\bibfnamefont{M.}~\bibnamefont{Ohzeki}},
  \bibinfo{journal}{Phys. Rev. E} \textbf{\bibinfo{volume}{80}},
  \bibinfo{pages}{011141} (\bibinfo{year}{2009}{\natexlab{b}}).

\end{thebibliography}

\end{document}